\documentclass[prd,aps,showpacs,superscriptaddress,nofootinbib,amsmath,amssymb]{revtex4}
\usepackage{graphics}
\usepackage{graphicx}
\usepackage{dcolumn}
\usepackage{bm}
\usepackage{lipsum}
\usepackage{multirow}
\usepackage{adjustbox}
\usepackage{textcomp}
\graphicspath{{./eps/}}
\begin{document}
\title{Weak quasielastic electroproduction of hyperons with polarization observables}

\author{F. \surname{Akbar}}
\affiliation{Department of Physics, Aligarh Muslim University, Aligarh-202002, India}

\author{M. Sajjad \surname{Athar}}
\email{sajathar@gmail.com}
\affiliation{Department of Physics, Aligarh Muslim University, Aligarh-202002, India}

\author{A. \surname{Fatima}}
\affiliation{Department of Physics, Aligarh Muslim University, Aligarh-202002, India}

\author{S. K. \surname{Singh}}
\affiliation{Department of Physics, Aligarh Muslim University, Aligarh-202002, India}
\def\a              {\alpha}
\def\b              {\beta}
\def\m              {\mu}
\def\n              {\nu}
\def\ti             {\tilde}
\def\e              {\varepsilon}
\def\r              {\rho}
\def\Gamma              {\varGamma}
\def\Sigma             {\varSigma}
\def\Lambda              {\varLambda}
\def\Delta              {\varDelta}

\begin{abstract}
With the availability of high luminosity electron beam at the accelerators, there is now the possibility of studying weak quasielastic hyperon 
production off the proton, \textit{i.e.} 
$e^-p \to \nu_e Y(Y=\Lambda,\Sigma^0)$, which will enable the determination of the nucleon-hyperon vector and axial-vector transition form factors 
at high $ Q^2$ in the strangeness sector and provide test of the
Cabibbo model, G-invariance, CVC, PCAC hypotheses and SU(3) symmetry. In this work, we have studied  
the total cross section, differential cross section as well as the longitudinal and perpendicular components of polarization of the final hyperons 
($ \Lambda$ and $\Sigma^0$ produced in these reactions) and presented numerical results for these observables and their sensitivity to the transition 
form factors. 
\end{abstract}
\maketitle
\section{Introduction}
The study of weak interaction processes at high energy and momentum transfers is done with the experiments performed using neutrino and 
antineutrino beams. The interpretation of these experiments to understand the QCD structure of nucleons and extract various parameters of 
weak interaction phenomenology suffers from the systematic uncertainties arising due to the lack of well-understood (anti)neutrino flux and the 
nuclear medium effects due to the presence of heavy nuclear targets used in the large volume detectors. An extensive discussion of these systematic 
uncertainties and theoretical attempts to model them exists in the literature~\cite{Katori:2016yel,Morfin:2012kn,Alvarez-Ruso:2014bla}. The presence 
of these systematic uncertainties can be eliminated if the monoenergetic beams of the charged lepton probes can be used with the proton as the 
target to study the weak interaction processes.

The study of such processes has been proposed almost 50 years back but has not been seriously pursued due to the very small cross sections making
it difficult to observe them experimentally~\cite{Fearing:1969nr}.
Theoretically, there have been very few calculations to study the weak interaction processes induced by the electrons and they have been 
limited to the study of the quasielastic processes in the $\Delta$S = 0 sector at very low energies from the nuclear targets relevant for the
astrophysical applications~\cite{Langanke:2002ab,Suzuki:2011zzc}. 
In the high energy region the study of weak inelastic excitations of $\Delta$ and $N^*$ in the $\Delta$S = 0 sector~\cite{AlvarezRuso:1997jr,
Hwang:1987sd,Nath:1979qe,Pollock:1998tz,Wang:2013kkc,Wang:2014guo,Jones:1979aj,Mukhopadhyay:1998mn,Hammer:1995dea} and 
 the quasielastic production of hyperons in the $|\Delta$S$|$ = 1 sector
 induced by the electrons on the protons have been studied in the recent past~\cite{Hwang:1988fp,Mintz:2004eu,Mintz:2002cj,Mintz:2001jc}.

 It has been shown in these studies that with the availability of high luminosity 
unpolarized and polarized electron beams as well as the unpolarized and polarized proton targets there is the possibility of performing electron 
scattering experiments to study the weak processes in the charged and neutral current sectors
at high energy and momentum transfers. Indeed, 
the polarized electrons have been used for the last many years to study the weak interaction processes which have been, however, limited to the neutral 
current sector. The study of the parity violating asymmetry in the scattering of polarized electrons from the proton targets 
has provided important information about the vector and axial-vector neutral current coupling of the
electrons to the quarks in the DIS processes~\cite{Prescott:1979dh,Cahn:1977uu,Brady:2011uy,Matsui:2005ns,Gorchtein:2011mz,Hall:2013hta,
Mantry:2010ki,Hobbs:2008mm}
and the $N-\Delta$ transition form factors in the inelastic processes~\cite{AlvarezRuso:1997jr,Hwang:1987sd,Nath:1979qe,Pollock:1998tz,Wang:2013kkc,
Wang:2014guo,Jones:1979aj,Mukhopadhyay:1998mn,Hammer:1995dea} as well as 
the presence of strangeness in the nucleon form factors in the quasielastic 
processes~\cite{Musolf:1993tb,GonzalezJimenez:2011fq,Armstrong:2012bi,Kumar:2013yoa,Beise:2004py}. 
However, no experimental attempts have been made to study the weak
processes induced by the high energy electrons in the charge current sector.

With the luminosity of $ 10^{39}-10^{40}/\rm{cm}^2/\rm{s}$ of the electron beam which may be presently 
available at JLab~\cite{Rode:2010zz,jlab} and MAMI~\cite{mainz}, it should be possible to study the weak production of $\Delta$ and hyperons 
($\Lambda$ and $\Sigma$). 

Although the hyperon production is suppressed as compared to the $\Delta$ production by a factor of $\rm{tan}^2 \theta_c$ where $\theta_c$ is the 
Cabibbo angle, it could be important in the energy region close to the threshold of $\Delta$ production where the $\Delta$ production cross section
is small due to the threshold effects. It would be, therefore, interesting to quantitatively study the kinematic region where the hyperon 
production becomes significant specially in the low energy region of electrons. At higher electron energies, the weak hyperon production is overwhelmed
 by the electromagnetic associated production of $\Lambda (\Sigma^0)$, \textit{i.e.} $e^- + p \longrightarrow e^- + \Lambda(\Sigma^0) + K^+ $, which happens at 
 the electron energies above the energy corresponding to the threshold of associated particle production processes.  
 However, the weak quasielastic production of $\Lambda$ is clearly distinguishable from the associated 
electroproduction of $\Lambda$ as it produces no electron in the final state but only the hadronic states of the nucleons and the pions 
through its decay 
products. 
Therefore, in this energy region the pion production without electrons is a clear signal of weak production of $\Lambda$ and $\Delta$ in the final 
state. However,
the electron induced weak production of pions can be seen even at lower energies corresponding to the threshold production of pions through the 
processes $e^- + p \longrightarrow \nu + n + \pi^+$ and $e^- + p \longrightarrow \nu + p + \pi^0$ which take place through the nonresonant processes
mediated by pions and nucleons as well as the contact term required by the gauge invariance. As the energy increases, the nonresonant and the 
resonant production along with $\Lambda$ production contribute to the weak pion production. The low energy weak production of pions in the threshold 
region is an important topic in itself and provides valuable information about the electroweak multipoles~\cite{Bernard:1993xh,Ch}. However, this has not been studied in the case
of threshold weak pion production induced by electrons and is beyond the scope of the present work.

In the case of quasielastic reactions whenever the $\Lambda$ and $\Sigma^0$ hyperons are produced by the charged current interaction, the 
observation of the differential cross section and the polarization of final hyperons can yield important information about the nucleon-hyperon transition form 
factors and enable the study of the applicability of Cabibbo model, G-invariance, T-invariance and SU(3) symmetry at high $Q^2$ in the strangeness 
sector. This would extend our understanding of the weak interaction phenomenology in the strangeness sector to high $Q^2$ which is presently available 
only at very low $Q^2$ from the study of semileptonic decays of hyperons~\cite{Cabibbo:2003cu,Gaillard:1984ny,Gazia}. The observation of hyperons
 in the final state through its decay 
products, \textit{i.e.} $\Lambda / \Sigma \longrightarrow N \pi$, and the structure of the angular distribution of pions will give information about the polarization 
of hyperons. The polarization observables of the hyperons produced in the quasielastic reactions induced by $\bar \nu_{\mu}$ 
are shown to be more sensitive to the weak axial form 
factors~\cite{Erriquez:1978pg,Alam:2014bya,Alam:2013cra,Pais:1971er,Marshak,LlewellynSmith:1971uhs,Akbar:2016awk}.

In view of the above discussion, we have studied in this paper
the total cross section, differential cross section and the polarization observables of the final hyperons produced in
\begin{equation}
 e^- + p \longrightarrow \Lambda (\Sigma^0) + \nu_e \label{hyp-rec}
\end{equation}
reactions and their sensitivity on the nucleon-hyperon transition form factors.
 
In section-\ref{hyperon}, the formalism to calculate the quasielastic weak hyperon production cross section and the expressions for the
differential cross section 
$(\frac{d\sigma}{dQ^2})$, longitudinal ($P_L(Q^2)$) and perpendicular ($P_P(Q^2)$) components of the hyperon polarization are given. In section-\ref{delta_sec}, we have given in brief 
the formalism to calculate the $\Delta^0$ production cross section for the electron on the proton target. In section-\ref{results}, we have 
presented the 
numerical results for the total cross section ($\sigma$), angular ($\frac{d\sigma}{d\Omega}$) and $Q^2$ ($\frac{d\sigma}{dQ^2}$) distributions 
and compared the results for the $Q^2$ distribution and $\sigma$ for the $\Lambda(\Sigma^0)$ productions with 
 the corresponding results for the $\Delta^0$ production. We have presented the numerical results for the longitudinal $P_L(Q^2)$ and perpendicular $P_P(Q^2)$ polarization components 
 of $\Lambda / \Sigma^0$ and discussed their sensitivity to the nucleon--hyperon transition from factors. All the numerical calculations have been 
 performed in the lab frame, \textit{i.e.}, assuming the nucleon to be at rest. Our findings are summarized in section-\ref{summary}.
  \section{Formalism}\label{formalism}
 \subsection{$e^- ~+~ p \rightarrow \nu_e ~+~ Y$ process}\label{hyperon}
 \begin{figure}
 \begin{center}
    \includegraphics[height=3cm,width=6cm]{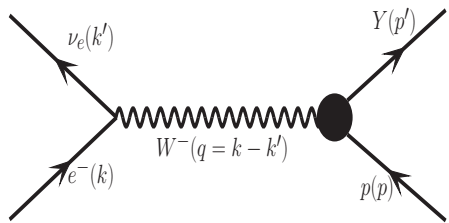}
  \caption{Feynman diagram  for the process $ e^-(k) + p(p) \rightarrow \nu_e(k^\prime) + Y(p^\prime)$, 
  where $Y$ stands for final hyperon. 
  The quantities in the bracket represent 
  four momentum of the corresponding particles.}\label{fyn_hyp}
   \end{center}
 \end{figure}
 \subsubsection{Cross section}
 The general expression of the differential cross section corresponding to the process presented in Fig.~\ref{fyn_hyp} may be written as
 \begin{equation}
 \label{crosv.eq}
 d\sigma=\frac{1}{(2\pi)^2}\frac{1}{4E_e^L m_N}\delta^4(k+p-k^\prime-p^\prime)
 \frac{d^3k^\prime}{2E_{k^\prime}}\frac{d^3p^\prime}{2E_{p^\prime}} \overline{\sum} \sum |{\cal{M}}|^2,
 \end{equation}
 where $E_e^L$ is the electron energy in the lab frame and the square of the transition matrix element is defined in terms of the leptonic 
 (${\cal L}_{\alpha \beta}$) and hadronic (${\cal J}^{\alpha \beta}$) tensors:
\begin{equation}\label{matrix}
  \overline{\sum} \sum |{\cal{M}}|^2 = \frac{G_F^2 \sin^2\theta_c}{2} \cal{J}^{\alpha \beta} \cal{L}_{\alpha \beta}. 
\end{equation}
  In the above expression, $G_F$ is the Fermi coupling constant. 
The hadronic and leptonic tensors are given by
\begin{eqnarray}\label{J}
\cal{J}_{\alpha \beta} &=& \frac{1}{2} \mathrm{Tr}\left[\Lambda(p')J_{\alpha}
 \Lambda(p)\tilde{J}_{\beta} \right] \\
 \label{L}\cal{L}^{\alpha \beta} &=& \frac{1}{2}\mathrm{Tr}\left[\gamma^{\alpha}(1-\gamma_{5})(k\!\!\!/+m_e)
\gamma^{\beta}(1-\gamma_{5})k'\!\!\!\!\!/~\right],
\end{eqnarray} 
with $\tilde{J}_{\beta} =\gamma^0 J^{\dagger}_{\beta} \gamma^0$ and $\Lambda(p)=p\!\!\!/+m_p$. 

 $J_{\a}$ is the hadronic current operator given by
  \begin{equation}\label{had}
  J_{\a} = V_{\a} - A_{\a},
 \end{equation}
 where
 \begin{eqnarray}\label{vec}
  V_{\a} = \gamma_\a f^{NY}_1(Q^2)+i\sigma_{\a\b} \frac{q^\b}{m_N+m_Y} f^{NY}_2(Q^2) + \frac{q_\a}{m_N+m_Y} f^{NY}_3(Q^2), 
 \end{eqnarray}
 and
 \begin{eqnarray}\label{axi}
  A_{\a} = \gamma_\a \gamma_5 g^{NY}_1(Q^2) + i \sigma_{\a\b}\gamma_5 \frac{q^\b}{m_N+m_Y} g^{NY}_2(Q^2) + \frac{q_\a} {m_N+m_Y} g^{NY}_3(Q^2) \gamma_5.
 \end{eqnarray}
 $m_N$ and $m_Y$ are the masses of the initial and final baryons, and $q_\mu (= k_\mu - k_\mu^{\prime}= p_\mu^{\prime}-p_\mu)$ is the four momentum 
 transfer with $ Q^2 = -q^2,~Q^2 \ge 0$.

Using the above definitions, the $Q^2$ distribution is written as
\begin{equation}\label{dsig}
 \frac{d\sigma}{dQ^2}=\frac{1}{64 \pi m_N^2 {E_e^L}^2} \overline{\sum} \sum |{\cal{M}}|^2.
\end{equation}
 In Eq. (\ref{dsig}), $|{\cal{M}}|^2$ is calculated using Eq.~(\ref{matrix}) assuming the absence of the second class currents and neglecting the 
 contribution from the pseudoscalar term due to the small mass 
 of the electron. The transition form factors $f^{NY}_i(Q^2)$ and $g^{NY}_i(Q^2)$ $(i=1-3)$, appearing in Eqs.~(\ref{vec}) and (\ref{axi}), 
 respectively, are 
 determined using the conservation of vector current (CVC), the partial conservation of axial current (PCAC), the principles of 
 T--invariance and G--invariance and the SU(3) symmetry.
\subsubsection{Form Factors}\label{sec:form}
The six form factors $f^{NY}_i(Q^2)$ and $g^{NY}_i(Q^2)$ ($i=1-3$) are determined using the following assumptions about the vector and
axial vector currents in the weak interactions:
\begin{enumerate}
\item[a)] The assumption of the T--invariance implies that all the form factors $f^{NY}_i(Q^2)$ and $g^{NY}_i(Q^2)$ given in Eqs.~(\ref{vec}) and 
(\ref{axi}), respectively, are real.
\item[b)] The assumption of the SU(3) symmetry of the weak hadronic currents implies that the vector and axial vector 
currents transform as an octet under the SU(3) group of transformations.

Since the initial and final baryons also belong to the octet representation, therefore, each form factor  $f^{NY}_i(Q^2)$ ($g^{NY}_i(Q^2)$) occurring 
in the matrix element of the vector (axial vector) current is written in terms of the two functions $D(Q^2)$ and $F(Q^2)$ corresponding to the symmetric 
octet($8^{S}$) and antisymmetric octet($8^{A}$) couplings of octets of vector (axial vector) currents. Specifically, we write
\begin{eqnarray}
 f^{NY}_i(Q^2) = a F_{i}^{V}(Q^2) + b D_{i}^{V}(Q^2)\label{coef1},\\
 g^{NY}_i(Q^2) = a F_{i}^{A}(Q^2) + b D_{i}^{A}(Q^2)\label{coef2},
\end{eqnarray}
where $a$ and $b$ are SU(3) Clebsch-Gordan coefficients given in Table-\ref{tab1}. $F_i^{V,A}(Q^2)$ and $D_i^{V,A}(Q^2)$ are the couplings corresponding 
to the antisymmetric and symmetric couplings of the two octets.

\item[c)] For the determination of the vector form factors we have assumed the CVC and 
SU(3) symmetry which lead to $f^{NY}_3(Q^2) = 0$. 
Further, the remaining two vector form factors \textit{viz.} $f^{NY}_1(Q^2)$ and $f^{NY}_2(Q^2)$ are determined 
in terms of the electromagnetic form factors of the nucleon, \textit{i.e.} 
 $f_{1}^{N}(Q^{2})$ and $f_{2}^{N}(Q^{2})$. 
 This is done by taking the matrix element of the electromagnetic current operator between the nucleon states and determining 
 $F_{i}^{V}(Q^2)$ and $D_{i}^{V}(Q^2)$ in terms of the electromagnetic form factors 
 $f^{N}_{i}(Q^2)$ $(i = 1, 2)$ of the nucleon.
 The  functions $F_{i}^{V}(Q^2)$ and $D_{i}^{V}(Q^2), \, (i=1,2)$ are thus expressed in terms of the 
 nucleon form factors $f_{1}^{p,n}(Q^{2})$ and $f_{2}^{p,n}(Q^{2})$ as 
\begin{eqnarray}\label{eq:fiv_div}
 F_i^V(Q^2) &=& f_i^p(Q^2) + \frac12 f_i^n (Q^2),  \\ 
 D_i^V(Q^2) &=& - \frac32 f_i^n (Q^2). 
\end{eqnarray}

The vector form factors $f^{NY}_i(Q^2), i=1,2$ are derived using Eq.~(\ref{coef1}) and are tabulated
in  Table-\ref{tab:formfac}.

\item[d)]  In the axial vector sector, the form factor $g^{NY}_{2}(Q^{2})$ vanishes due to G--invariance, T--invariance and SU(3) symmetry. 
The axial vector form factor $g^{NY}_{1}(Q^{2})$ is determined from Eq.~(\ref{coef2}).
We write the  axial vector form factor $g^{NY}_1(Q^2)$ in terms of 
two functions $F_1^A(Q^2)$ and $D_1^A(Q^2)$. Using Table-\ref{tab1} 
for the coefficients $a$ and $b$, we find 
\begin{eqnarray}
  g_1^{p\Lambda}(Q^2)&=&\sqrt{\frac16}\left(3F^A_1(Q^2)+D^A_1(Q^2)\right), \nonumber\\
  g_1^{p\Sigma^0}(Q^2)&=& \sqrt{\frac12} \left[ D^A_1(Q^2)-F^A_1(Q^2) \right],
\end{eqnarray}\label{Eq:xdep}
which are rewritten in terms of the axial vector form factor $g_A^{n  p}(Q^2)(= F_1^A(Q^2)+D_1^A(Q^2))$ 
for the $n-p$ transition and are given in Table-\ref{tab:formfac} with $x(Q^2)$ defined as 
\begin{equation}
x(Q^2)=\frac{F^A_1(Q^2)}{F^A_1(Q^2)+D^A_1(Q^2)}.
\end{equation} 

We further assume that $F^A_1(Q^2)$ and $D^A_1(Q^2)$ have the same $Q^2$ dependence, such that 
$x(Q^2)$ becomes a constant and is given by 
$x=\frac{F^A_1(0)}{F^A_1(0)+D_1^A(0)}$. For $g_A^{np}(Q^2)$, a dipole parameterization has been used 
\begin{equation}
 g_A^{np}(Q^2)=g_A(0)\left(1+\frac{Q^2}{M_A^2}\right)^{-2},
\end{equation}
where $M_A$ is the axial dipole mass and 
 for the numerical calculations we have used $M_A=1.026$ GeV~\cite{Bernard:2001rs}.
The axial charge $g_A(0) = 1.2723$ and $x = 0.364$~\cite{Cabibbo:2003cu} are
 determined from the experimental data  on the $\beta$ decay of neutron and the semileptonic decay of hyperons.
\end{enumerate}

\begin{table}\centering
 \begin{tabular}{|c|c|c|c|c}\hline
 & ~$p \to \Lambda$~ &  ~$ p \to \Sigma^{0}$~&~$p \to n $ \\ \hline \hline 
~~$a$~~~& ~~~$-\sqrt{\frac{3}{2}}$~~~ & ~~~~$-\frac{1}{\sqrt{2}}$~~~~ & ~~~~~1~~~~~ \\ \hline 
 ~~$b$~~~& ~~~$-\sqrt{\frac{1}{6}}$~~~ & ~~~~$\frac{1}{\sqrt{2}}$~~~~ & ~~~~~1~~~~~\\ \hline 
 \end{tabular}
\caption{Values of the coefficients $a$ and $b$ given in Eqs.~(\ref{coef1}) and (\ref{coef2}).}
\label{tab1}
\end{table}

\begin{table}[h!]
 \begin{center}
\begin{adjustbox}{max width=\textwidth} 
\begin{tabular}{|c|c|c|}  \hline 
&$e^- p \rightarrow \nu_e \Lambda$&$e^- p \rightarrow \nu_e \Sigma^0$\\ \hline  \hline            
 $f_1^{NY}(Q^2)$&$ -\sqrt{\frac{3}{2}}~f_1^p(Q^2)$&$-\frac{1}{\sqrt2}\left[f_1^p(Q^2) + 2 f_1^n(Q^2) \right]$\\ \hline
$f_2^{NY}(Q^2)$&$-\sqrt{\frac{3}{2}}~f_2^p(Q^2)$&$-\frac{1}{\sqrt2}\left[f_2^p(Q^2) + 2 f_2^n(Q^2) \right]$\\ \hline
$g_1^{NY}(Q^2)$&$-\frac{1}{\sqrt{6}}(1+2x) g_A(Q^2)$&$\frac{1}{\sqrt2}(1-2x)g_A(Q^2)$\\ \hline 
  \end{tabular}
  \end{adjustbox}
\end{center}
\caption{Vector and axial vector from factors for $e^-(k) + p(p)\rightarrow \nu_e(k^\prime) + Y(p^\prime)$ processes.}
 \label{tab:formfac}
\end{table}

\subsubsection{Polarization of hyperons}
Using the covariant density matrix formalism, the polarization 4-vector($\xi^\tau$) of the final hyperon produced
in reaction~(\ref{hyp-rec}) is written as~\cite{Bilekny}
\begin{equation}\label{polar}
\xi^{\tau}=\frac{\mathrm{Tr}[\gamma^{\tau}\gamma_{5}~\rho_{f}(p^\prime)]}
{\mathrm{Tr}[\rho_{f}(p^\prime)]},
\end{equation}
where the final spin density matrix $\rho_f(p^\prime)$ is given by 
\begin{equation}\label{polar1}
 \rho_{f}(p^\prime)= {\cal L}^{\alpha \beta}  \Lambda(p')J_{\alpha} \Lambda(p)\tilde{J}_{\beta} 
\Lambda(p').
\end{equation} 
Using the following relations~\cite{Bilenky:2013fra,Bilenky:2013iua}:
\begin{equation}\label{polar3}
\Lambda(p')\gamma^{\tau}\gamma_{5}\Lambda(p')=2m_Y\left(g^{\tau\sigma}-
\frac{p'^{\tau}p'^{\sigma}}{m_Y^{2}}\right)\Lambda(p')\gamma_{\sigma}
\gamma_{5}
\end{equation}
and
\begin{equation}
 \Lambda(p^\prime)\Lambda(p^\prime) = 2m_Y \Lambda(p^\prime),
\end{equation}
 $\xi^\tau$ defined in Eq.~(\ref{polar}) may be rewritten as
\begin{equation}\label{polar4}
\xi^{\tau}=\left( g^{\tau\sigma}-\frac{p'^{\tau}p'^{\sigma}}{m_Y^2}\right)
\frac{  {\cal L}^{\alpha \beta}  \mathrm{Tr}
\left[\gamma_{\sigma}\gamma_{5}\Lambda(p')J_{\alpha} \Lambda(p)\tilde{J}_{\beta} \right]}
{ {\cal L}^{\alpha \beta} \mathrm{Tr}\left[\Lambda(p')J_{\alpha} \Lambda(p)\tilde{J}_{\beta} \right]}.
\end{equation}
Note that in Eq.~(\ref{polar4}), $\xi^\tau$ is manifestly orthogonal to $p^{\prime \tau}$, \textit{i.e.} $p^\prime \cdot \xi=0$. Moreover, the 
denominator
is directly related to the 
differential cross section given in Eq.~(\ref{dsig}).

With ${\cal J}^{\a \b}$ and ${\cal L}_{\a \b}$ given in Eqs.~(\ref{J}) and (\ref{L}), respectively, an expression for $\xi^\tau$ is obtained. In the 
lab frame where the initial nucleon 
is at rest, the polarization vector $\bm{\xi}$ is calculated to be a function of 3-momenta of incoming electron $(\textbf{\textit{k}})$ and outgoing 
hyperon $(\textit{\textbf{p}}^{\prime})$, and is given as  
\begin{equation}\label{3pol}
 \bm{\xi} =\left[{\bm k} ~\alpha(Q^2) + {\bm p}\,^{\prime} \beta(Q^2) \right], 
\end{equation}
where the expressions of $\alpha(Q^2)$ and $\beta(Q^2)$ are given in the
appendix.

From Eq.~(\ref{3pol}), it follows that the polarization vector lies in the scattering plane defined by ${\bm k}$ and $\bm{p}\,^\prime$, and there is 
no component of polarization in a direction orthogonal to the scattering 
plane. This is a consequence of T--invariance which makes the transverse polarization in a direction perpendicular to the reaction plane
vanish~\cite{Pais:1971er,LlewellynSmith:1971uhs}. We now expand the polarization vector $\bm{\xi}$ along the two orthogonal directions, ${\bm e}_L$ and 
${\bm e}_P$ in the reaction plane corresponding to the longitudinal and perpendicular 
directions, to the momentum of hyperon \textit{i.e.}
\begin{equation}\label{vectors}
\bm{ e}_{L}=\frac{\bm{ p'}}{|\bm{ p'}|},\qquad
\bm{ e}_{P}=\bm{ e}_{L}\times \bm{ e}_T,\qquad   {\rm where}~~~~~ \bm{e}_T=\frac{\bm{ p'}\times \bm{ k}}{|\bm{ p'}\times \bm{ k}|},
\end{equation}
and write
 \begin{equation}\label{polarLab}
\bm{\xi}=\xi_{P} \bm{e}_{P}+\xi_{L} \bm{e}_{L},
\end{equation}
such that the longitudinal and perpendicular components of the polarization vector ($\bm{\xi}$) in the lab frame are given by
\begin{equation}\label{PL}
 \xi_L(Q^2)=\bm{\xi} \cdot \bm{e}_L,\qquad \xi_P(Q^2)= \bm{\xi} \cdot \bm{e}_P.
\end{equation}
From Eq.~(\ref{PL}), the longitudinal $P_L(Q^2)$ and perpendicular $P_P(Q^2)$ components of the polarization vector defined in the rest frame 
of the initial nucleon are given by ~\cite{Bilenky:2013fra,Bilenky:2013iua}
\begin{equation}\label{PL1}
 P_L(Q^2)=\frac{m_Y}{E_{p^\prime}} \xi_L(Q^2), \qquad P_P(Q^2)=\xi_P(Q^2),
\end{equation}
where $\frac{m_Y}{E_{p^\prime}}$ is the Lorentz boost factor along ${\bm p} ^\prime$.
With the help of Eqs.~(\ref{3pol}), (\ref{vectors}), (\ref{PL}) and (\ref{PL1}), the longitudinal component $P_L(Q^2)$ is calculated to be
\begin{equation}\label{sl}
P_L(Q^2) = \frac{m_Y}{E_{p^\prime}}\left(\frac{\alpha(Q^2) {\bm{k}}\cdot {\bm{p}}\,' + \beta(Q^2)|{\bm {p}}\,'|^2}{|{\bm {p}}\,'|~{\cal J}^{\alpha \beta} {\cal L}_{\alpha \beta} }\right),
\end{equation}
where $E_{p^{\prime}} = \sqrt{|{{\bm p}^\prime}|^2+m_Y^2}$. 
Similarly, the perpendicular component $P_P(Q^2)$  of the polarization 3-vector is given as
\begin{equation}\label{st}
 P_P(Q^2) = \frac{({\bm{k}}\cdot {\bm{p}^ \prime})^2 - |{\bm {k}}|^2|{\bm {p}}\,'|^2}{|{\bm {p}}\,'||{\bm {p'}}\times {\bm{k}}|~{\cal J}^{\alpha \beta} {\cal L}_{\alpha \beta}} \alpha(Q^2).
\end{equation}
\begin{figure}
 \includegraphics[height=6cm,width=10cm] {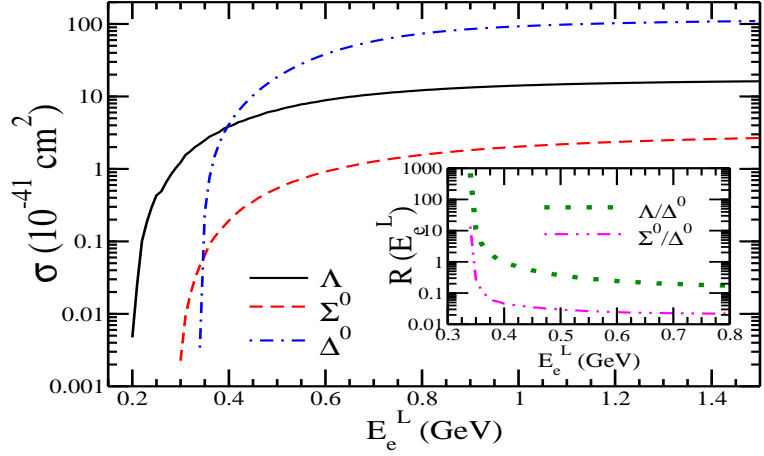}
 \caption{$\sigma$ \textit{vs.} $E_e^L$ for $\Lambda$ (solid line), $\Sigma^0$ (dashed line) and $\Delta^0$ (dashed-dotted line) productions.
  In the inset the results are presented for the ratio $R(E_e^L)=\frac{\sigma_Y}{\sigma_{\Delta^0}}$ \textit{vs.} $E_e^L$, for 
  Y=$\Lambda$ (dotted line), $\Sigma^0$ (dashed double-dotted line).}\label{sigma1}
\end{figure}
\begin{figure*}
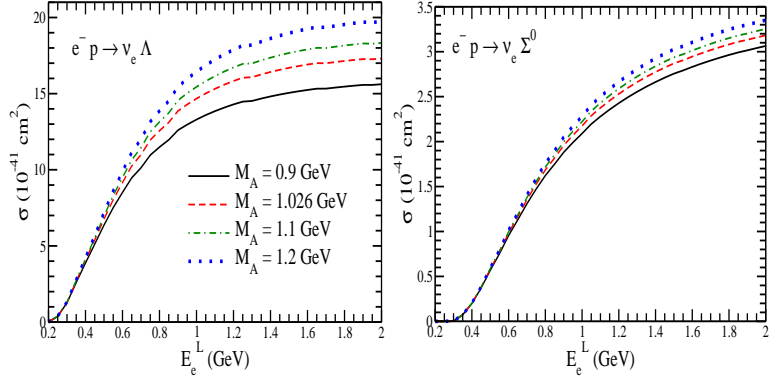

\begin{center}
  \includegraphics[height=5cm,width=5cm]{Ma_variation_lam_sigma_mintz_2GeV.eps}
   \includegraphics[height=5cm,width=5cm]{Ma_variation_sig0_sigma_mintz_2GeV.eps}
   \end{center}
  \caption{{$\sigma$ \textit{vs.} $E_e^L$ for $e^- p \to \nu_e \Lambda$ (left panel) and $e^- p \to \nu_e \Sigma^0$ (right panel) processes.
  The results are presented for different values of $M_A$ \textit{viz.} 0.9 GeV (solid line), 
  1.026 GeV (dashed line), 1.1 GeV (dash-dotted line) and 1.2 GeV (dotted line).}}\label{ma_sig}
\end{figure*}
\subsection{$e^- ~+~ p \rightarrow \nu_e ~+~ \Delta^0$ process}\label{delta_sec}
In order to compare the cross section for the hyperon production with the cross section for the $\Delta^0$ production, produced in the 
reaction
  \begin{equation}\label{delta}
    e^-(k) ~+~ p(p) \rightarrow \nu_e(k^\prime) ~+~ \Delta^0(p^\prime),
  \end{equation}
   we give the expression for the differential cross section for the $\Delta^0$ production as~\cite{AlvarezRuso:1997jr}

\begin{eqnarray} \label{delta_cross_section}
\frac {d \sigma}{d Q^2} = \frac{1}{16\pi^2} \int {d |{\bm{p^\prime}}|}\frac{{|{{\bm p} ^\prime}|}}{E_e^2 E_{\nu_e}} 
\frac{\frac{\Gamma_\Delta(W)}{2}}{(W-M_\Delta)^2+\frac{\Gamma_\Delta^2(W)}{4}} \overline{\sum} \sum |{\cal{M}}|^2. 
\end{eqnarray}
In the above expression $ \overline{\sum} \sum |{\cal{M}}|^2= \frac{G_F^2}{2} \cos^2\theta_c\, \cal{L}_{\mu\nu} $ $J^{\mu\nu}$, where the 
leptonic tensor $\cal{L}_{\mu\nu}$ is given in Eq.~(\ref{L}) and the hadronic tensor 
 $J^{\mu\nu}=\frac{1}{2} Tr\left[ \frac{(\not p + m_N)}{2 m_N}{\tilde{\mathcal O}}^{\alpha\mu } {\it P}_{\alpha \beta}
      {\mathcal O}^{\beta\nu} \right]$. The hadronic tensor is obtained by using the expression for 
      the hadronic current $j^{\mu}$ as
\begin{equation}
<\Delta(p^{\prime})|j^\mu|p(p)>= \bar \Psi_{\beta}(p^{\prime}){ \mathcal O}^{\beta \mu} u( p).
\end{equation}
In the above expression $u(p)$ is the Dirac spinor for the proton and  $\Psi_\beta$ is a Rarita-Schwinger field for spin-$\frac{3}{2}$ particle.
 $\mathcal O^{\beta \alpha}$ is the $N-\varDelta$ transition vertex, which is described in terms of the vector($C_{i}^{V}(q^2)$)
 and the axial vector($C_{i}^{A}(q^2)$) transition form factors with $\mathcal O^{\beta \alpha}={\mathcal O}_V^{\beta \alpha}+{\mathcal O}_A^{\beta \alpha}$, 
 which are given by
\begin{eqnarray}\label{vec_tra_current}
{\mathcal O}_V^{\beta \alpha}=\left(\frac{C_{3}^V(q^2)}{m_N}(g^{\alpha \beta}\not\! q-q^{\beta}\gamma^{\alpha})
+\frac{C_{4}^V(q^2)}{m_N^2}(g^{\alpha \beta} q \cdot p^{\prime}-q^{\beta}p^{\prime\alpha}) + 
\frac{C_{5}^V(q^2)}{m_N^2}(g^{\alpha \beta}q \cdot p-q^{\beta}p^{\alpha})\right)\gamma_{5}
\end{eqnarray}
and
\begin{eqnarray}\label{ax_tra_current}
{\mathcal O}_A^{\beta \alpha}=\left(\frac{C_{4}^{A}(q^2)}{m_N}(g^{\alpha \beta}{\not\! q} -q^{\beta}\gamma^{\alpha})
+C_{5}^{A}(q^2)g^{\alpha \beta}+\frac{C_{6}^{A}(q^2)}{m_N^2}q^{\beta}q^{\alpha}\right).
\end{eqnarray}
For the numerical calculations, we have taken the parameterization of Lalakulich \textit{et al.}~\cite{Lalakulich:2006sw} for $C_i^V(Q^2)$ and $C_i^A(Q^2)$:
\begin{equation}\label{civ_lala}
C_i^V(Q^2)=C_i^V(0)~\left(1+\frac{Q^2}{M_V^2}\right)^{-2}~{\cal{D}}_i,~~~i=3,4,5,
\end{equation}
with $C_3^V(0)=2.13$, $C_4^V(0)=-1.51$ and $C_5^V(0)=0.48$,
\begin{eqnarray}\label{di} 
{\cal{D}}_{3,4}&=&\left(1+\frac{Q^2}{4M_V^2}\right)^{-1}~~~ \mbox{and}\nonumber\\
{\cal{D}}_{5}&=&\left(1+\frac{Q^2}{0.776M_V^2}\right)^{-1};~~ M_V=0.84~\rm{GeV}
\end{eqnarray} 
and 
\begin{eqnarray}\label{cia_lala}
C_i^A(Q^2)&=&C_i^A(0)\left(1+\frac{Q^2}{M_A^2}\right)^{-2}\left(1+\frac{Q^2}{3M_A^2}\right)^{-1};\nonumber\\
M_A &=& 1.026~\rm{GeV}\\ \nonumber
\end{eqnarray}
for $i=3,4,5,6$ with $C_3^A(0)=0$, $C_4^A(0)=-0.25~C_5^A(0)$, $C_5^A(0)=-1.2$ and $C_6^A(0)=\frac{m_N^2}{(m^2_{\pi}+Q^2)}C_5^A(0)$.

\begin{figure*}
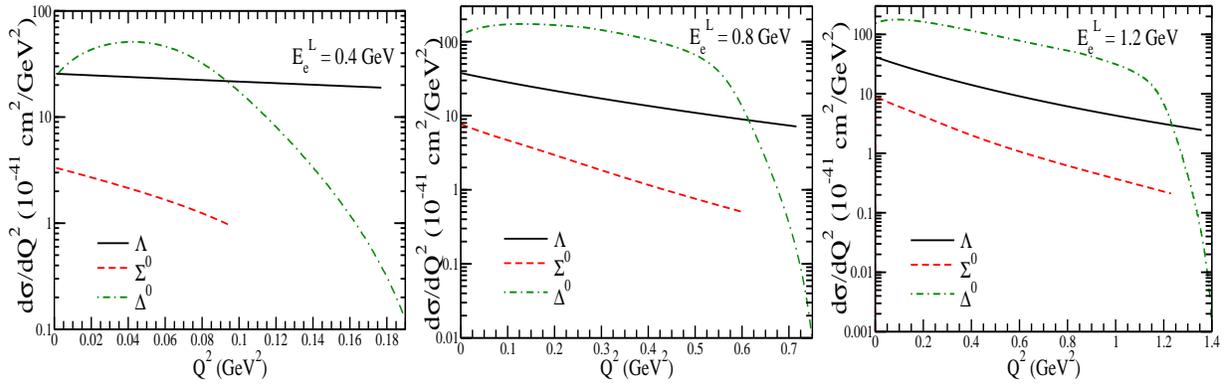

\begin{center}
 \includegraphics[height=5cm,width=5.3cm]{dsigma_dq2_comparison_ratio_elep_400Mev.eps}
 \includegraphics[height=5cm,width=5.3cm]{dsigma_dq2_comparison_ratio_elep_800Mev.eps}
 \includegraphics[height=5cm,width=5.3cm]{dsigma_dq2_comparison_ratio_elep_1200MeV.eps}
 \end{center}
  \caption{$\frac{d\sigma}{dQ^2}$ \textit{vs.} $Q^2$ at $E_e^L$ = 0.4 GeV (left panel), $E_e^L$ = 0.8 GeV (central panel) and $E_e^L$ = 1.2 GeV (right panel).
  The results are shown for $\Lambda$ (solid line), $\Sigma^0$ (dashed line) and $\Delta^0$ (dash-dotted line) productions.
}\label{dsigma1}
\end{figure*}
  \begin{figure*}
  \begin{center}
  \includegraphics[height=6cm,width=10cm]{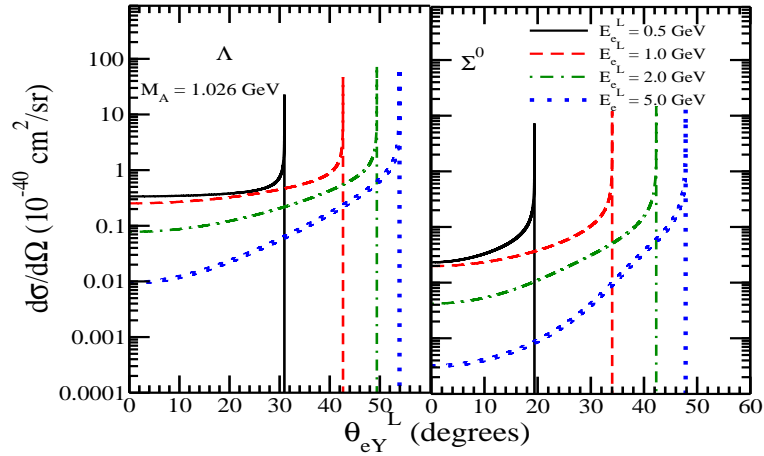}
  \end{center}
 \caption{{Hyperon angular distribution ($\frac{d\sigma}{d\Omega}$) \textit{vs.} outgoing $Y$ laboratory angle ($\theta_{eY}^L$)
 for the processes $e^- p \to \nu_e \Lambda$ (left panel) and $e^- p \to \nu_e \Sigma^0$ (right panel)
 with $M_A$=1.026 GeV at different electron energies $E_e^L$= 0.5 GeV (solid line), 1 GeV (dashed line),
 2 GeV (dash-dotted line) and 5 GeV (dotted line).}}\label{dom1}
\end{figure*}
\begin{figure*}
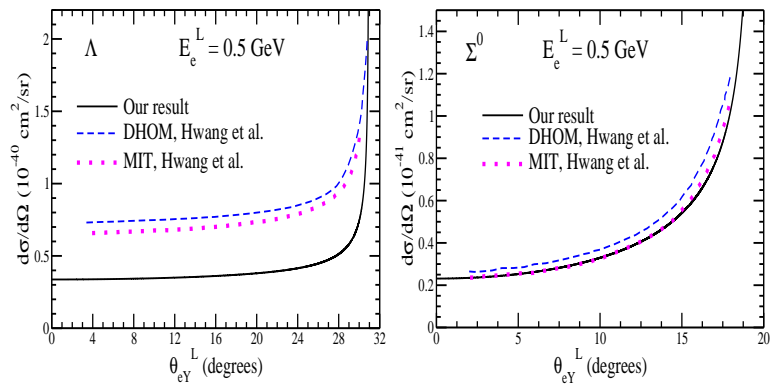

\begin{center}
 \includegraphics[height=5cm,width=5cm]{Lambda_hwang_comparison.eps}
 \includegraphics[height=5cm,width=5cm]{Sigma_hwang_comparison.eps}
 \end{center}
 \caption{{Hyperon angular distribution ($\frac{d\sigma}{d\Omega}$) \textit{vs.} outgoing $Y$ laboratory angle ($\theta_{eY}^L$) 
 for the processes $e^- p \to \nu_e \Lambda$ (left panel)
 and $e^- p \to \nu_e \Sigma^0$ (right panel) at $E_e^L$ = 0.5 GeV with $M_A = $ 1.026 GeV (solid line). 
 The results are compared with the results obtained by Hwang \textit{et al.}~\cite{Hwang:1988fp} using two different models, DHOM (dashed line) and MIT model (dotted line).}}\label{hwang_com}
\end{figure*}
\begin{figure*}
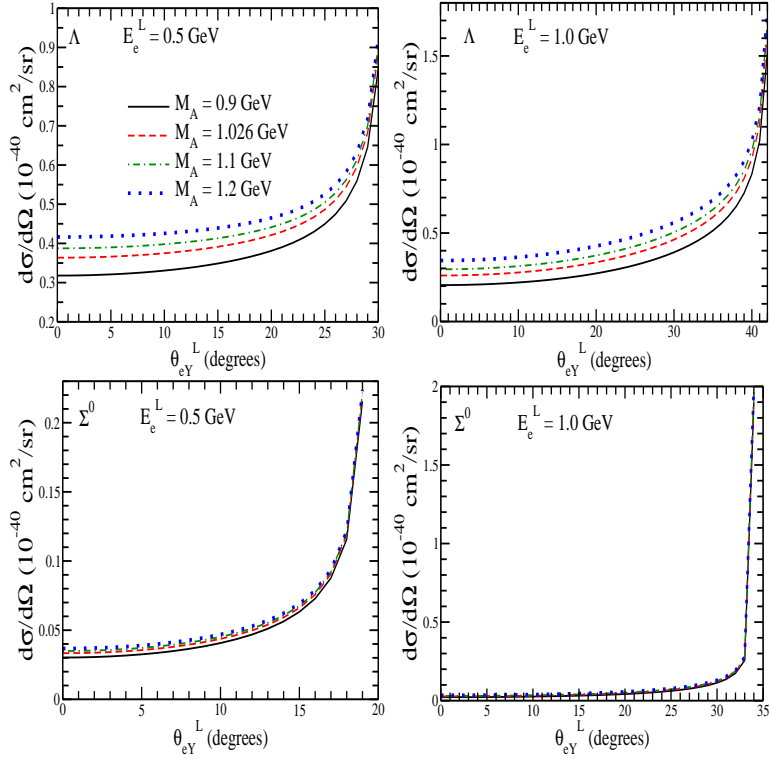

\begin{center}
 \includegraphics[height=5cm,width=5cm]{Ma_variation_lam_500MeV.eps}
 \includegraphics[height=5cm,width=5cm]{Ma_variation_lam_1GeV.eps}\\
 \includegraphics[height=5cm,width=5cm]{Ma_variation_sig0_500MeV.eps}
 \includegraphics[height=5cm,width=5cm]{Ma_variation_sig0_1GeV.eps}
 \end{center}
 \caption{{$\frac{d\sigma}{d\Omega}$ \textit{vs.} outgoing $\theta_{eY}^L$ for the processes $e^- p \to \nu_e \Lambda$ (upper panel)
 and $e^- p \to \nu_e \Sigma^0$ (lower panel) at $E_e^L$ = 0.5 GeV (left panel), $E_e^L$ = 1 GeV (right panel). The results are presented for different
 values of $M_A$ \textit{viz.} 0.9 GeV (solid line), 1.026 GeV (dashed line), 1.1 GeV (dash-dotted line) and 1.2 GeV (dotted line).}}\label{ma_dom}
\end{figure*}
${\it P}_{\alpha \beta}$ is the spin-3/2 projection operator given by
\begin{eqnarray}\label{propagator}
{\it P}_{\alpha \beta} = -(\frac{\not\! p^ \prime +M_{\Delta}}{2 M_{\Delta}}) 
\left(g_{\alpha \beta}-\frac{2}{3} \frac{{p^\prime}_{\alpha}{p^\prime}_{\beta}}{M_{\Delta}^2}
+ \frac{1}{3}\frac{{p^\prime}_{\alpha} \gamma_{\beta}-
{p^\prime}_{\alpha} \gamma_{\beta}}{M_{\Delta}}-\frac{1}{3}\gamma_{\alpha}\gamma_{\beta}\right),
\end{eqnarray}
and the delta decay width $\Gamma$ is taken as the energy dependent $P$-wave decay width given by
\begin{equation}
\Gamma_\Delta(W)=\frac{1}{6 \pi}\left(\frac{f_{\pi N \Delta}}{m_{\pi}}\right)^2 \frac{M_{\Delta}}{W}|\bm q_{cm}|^3,
\end{equation}
 where the $N-\Delta$ coupling constant $f_{\pi N \Delta}=2.127$, $m_{\pi}$ is the pion mass, $|\bm q_{cm}|$ is the pion momentum in the rest 
 frame of the resonance and is given by
\[|{\bm q_{cm}}|=\frac{\sqrt{(W^2-m_{\pi}^2-M_N^2)^2 -4 m_{\pi}^2M_N^2}}{2W},\] 
 with W [$(m_N+m_\pi)\le W < 1.4~\rm{GeV}$] as the center-of-mass energy.
\begin{figure*}
\begin{center}
 \includegraphics[height=5cm,width=5cm]{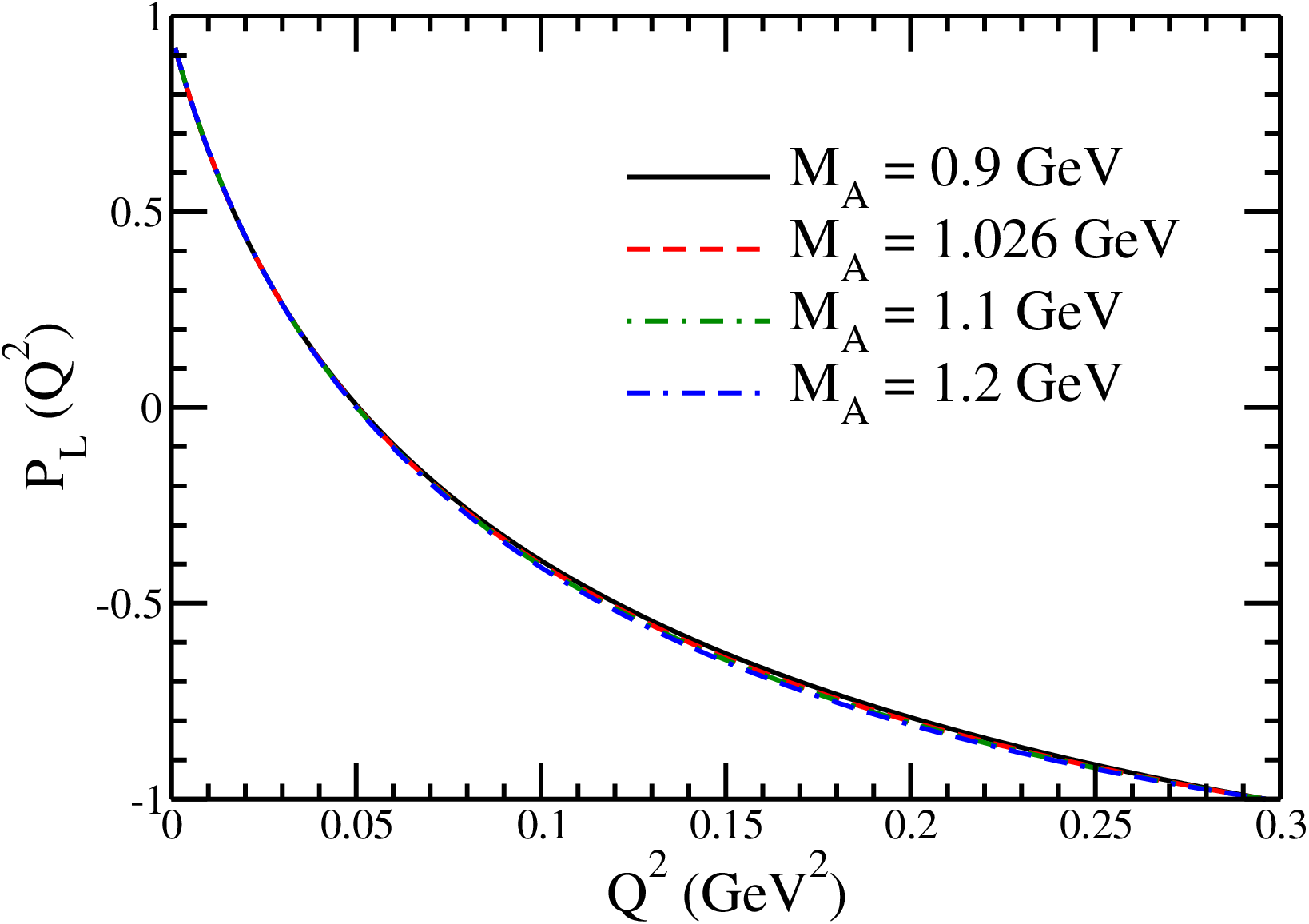}
 \includegraphics[height=5cm,width=5cm]{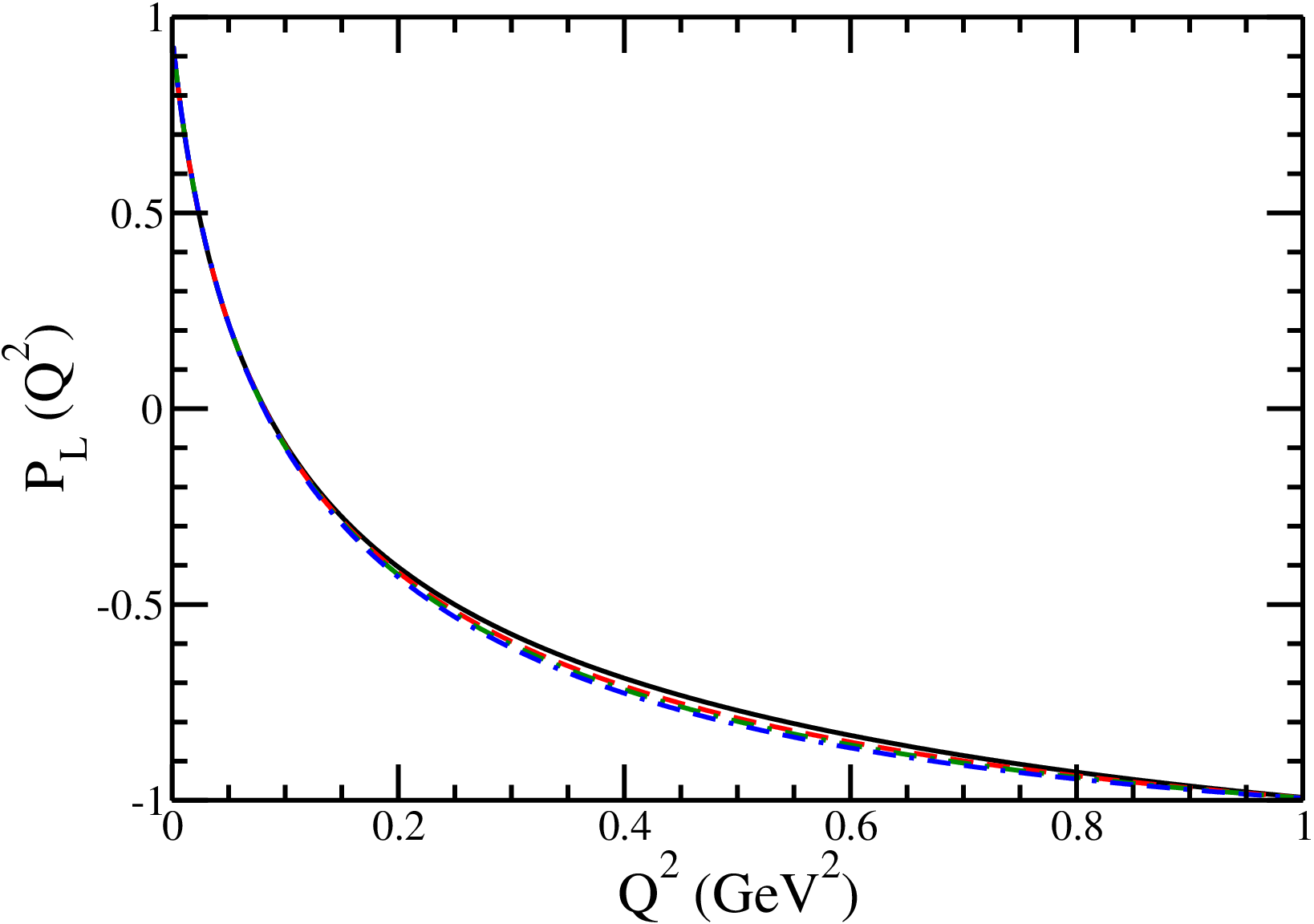}
 \includegraphics[height=5cm,width=5cm]{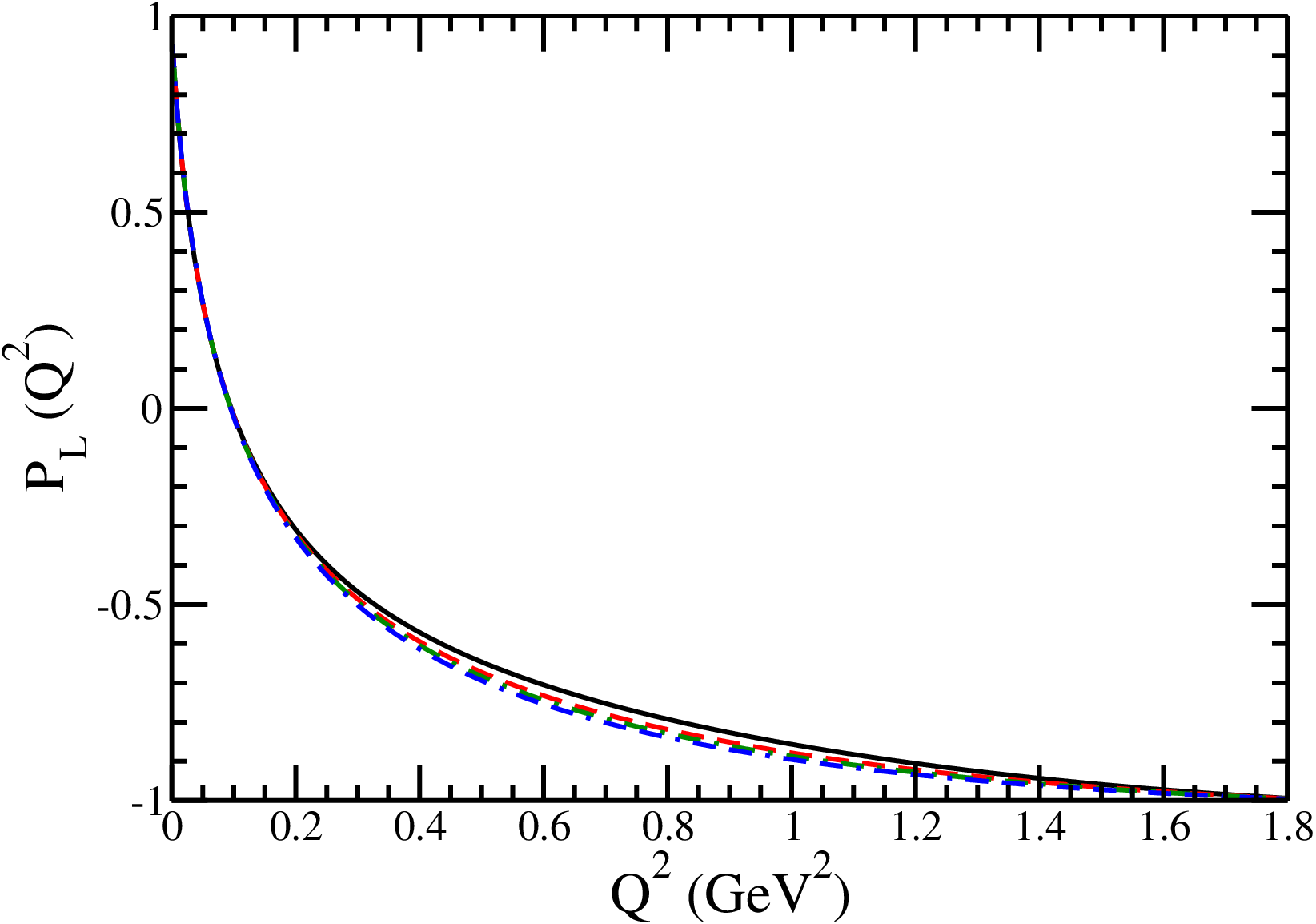} \\
 \includegraphics[height=5cm,width=5cm]{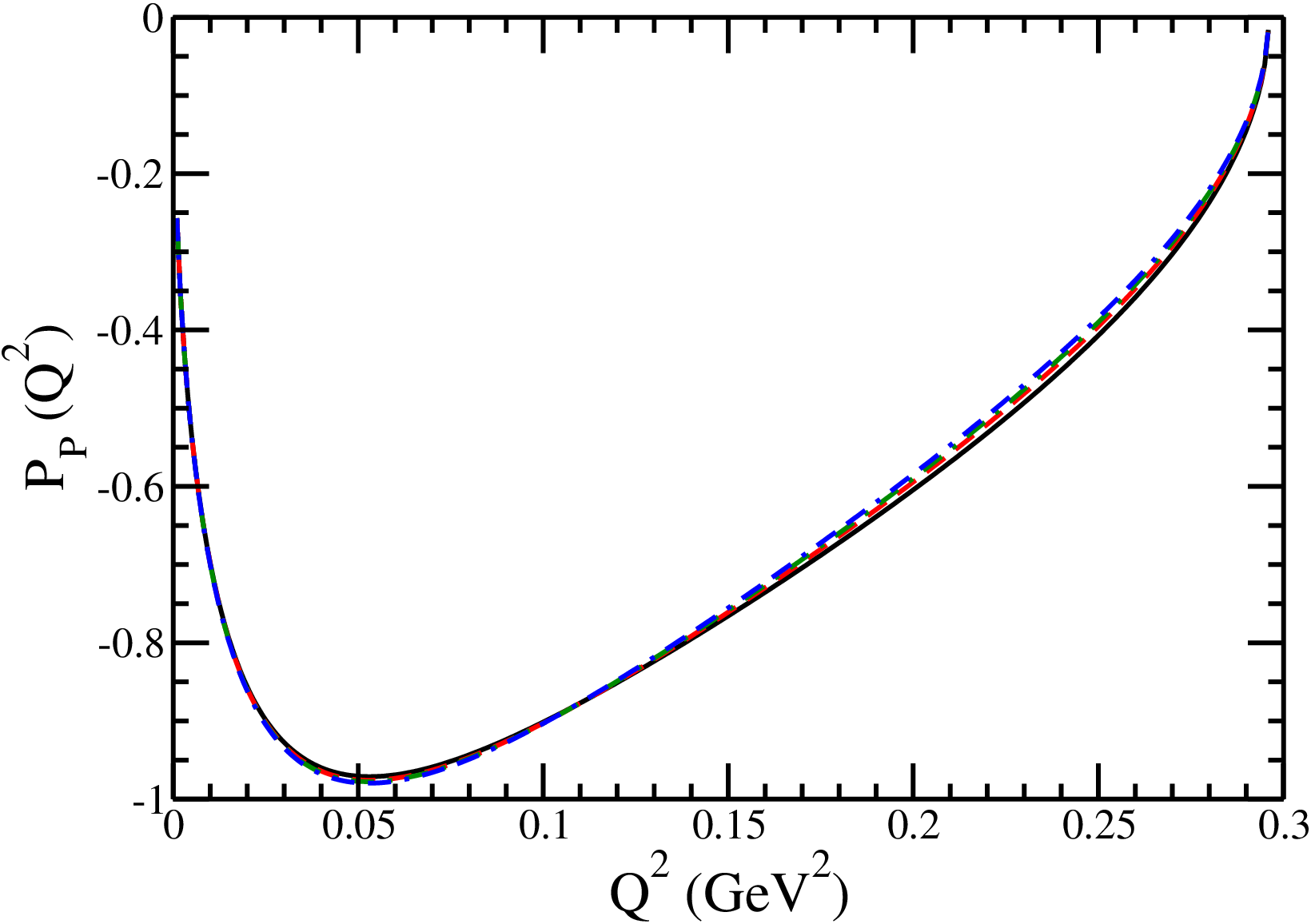}
 \includegraphics[height=5cm,width=5cm]{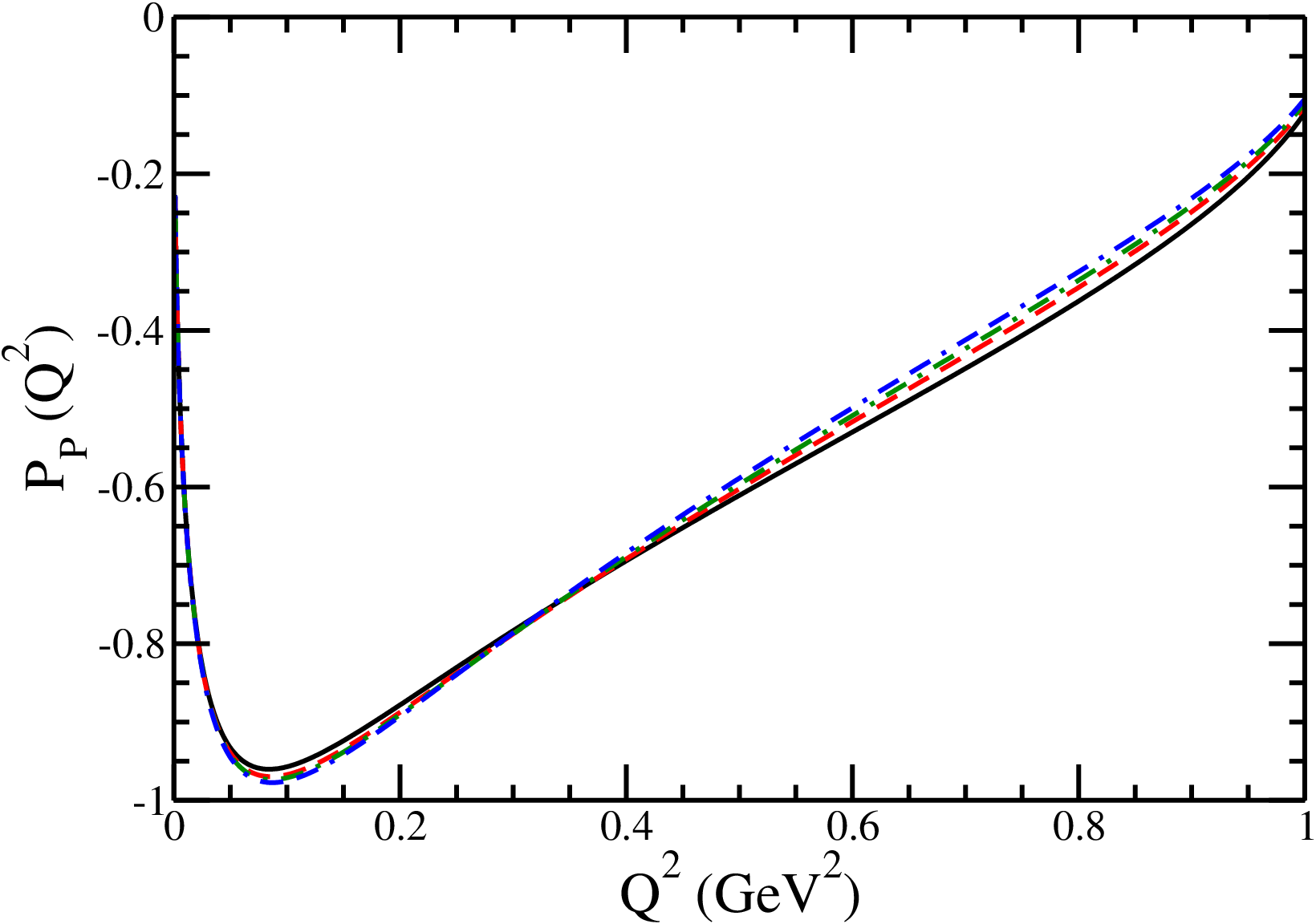}
 \includegraphics[height=5cm,width=5cm]{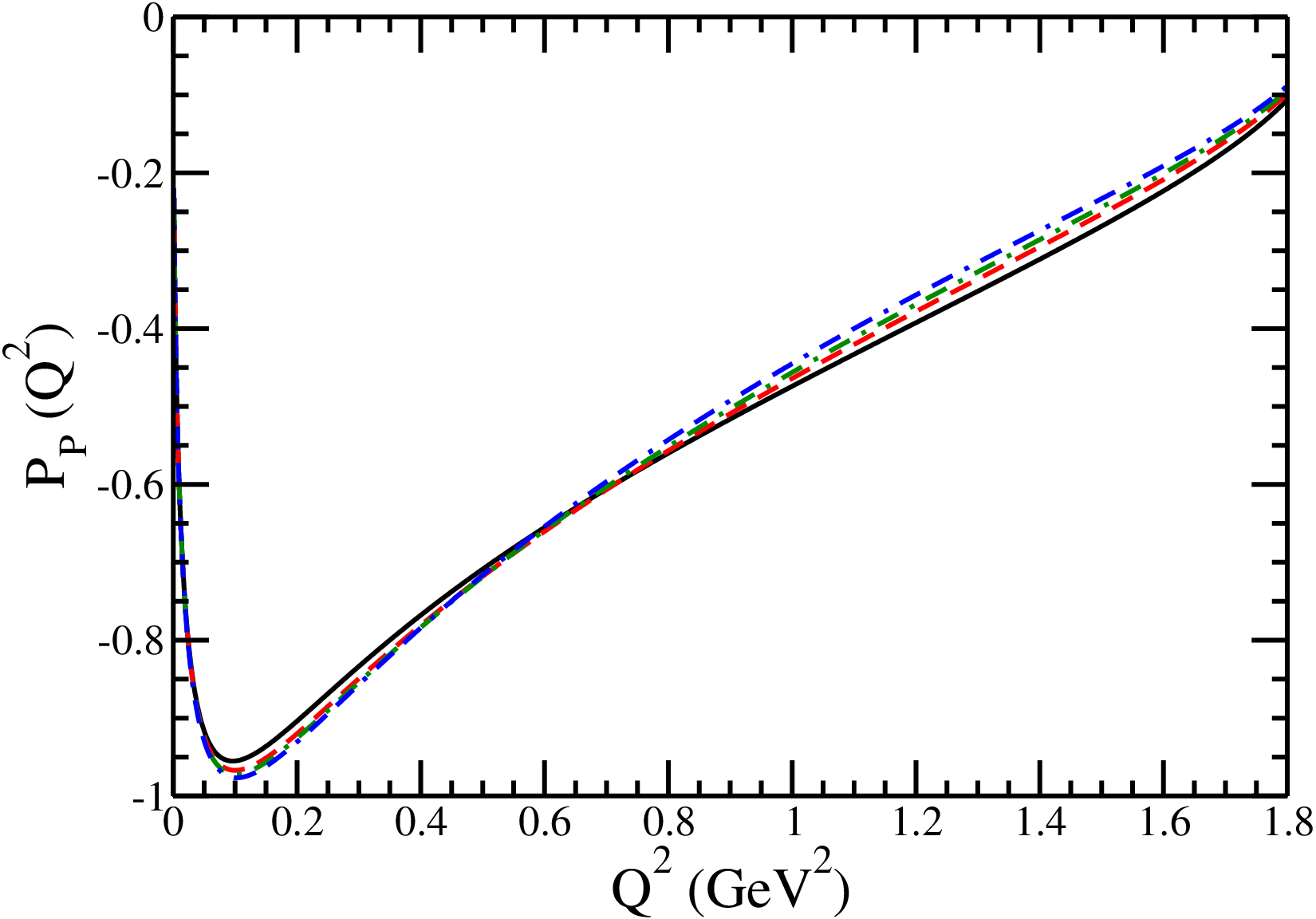}
 \end{center}
 \caption{{$P_L(Q^2)$ and $P_P(Q^2)$ \textit{vs.} $Q^2$ for the process $e^- p \to \nu_e \Lambda$ 
 at $E_e^L$ = 0.5 GeV (left panel), $E_e^L$ = 1 GeV (central panel) and $E_e^L$ = 1.5 GeV (right panel). The results are presented for different
 values of $M_A$ \textit{viz.} 0.9 GeV (solid line), 1.026 GeV (dashed line), 1.1 GeV (dash-dotted line) and 1.2 GeV (dash--double-dotted line).}}\label{ma_var_pol_lam}
\end{figure*}
\begin{figure*}
 \begin{center}

 \includegraphics[height=5cm,width=5cm]{Pl_Ma_variation_elep_500MeV_sigma0.eps}
 \includegraphics[height=5cm,width=5cm]{Pl_Ma_variation_elep_1GeV_sigma0.eps}
 \includegraphics[height=5cm,width=5cm]{Pl_Ma_variation_elep_1500MeV_sigma0.eps} \\
 \includegraphics[height=5cm,width=5cm]{Pp_Ma_variation_elep_500MeV_sigma0.eps}
 \includegraphics[height=5cm,width=5cm]{Pp_Ma_variation_elep_1GeV_sigma0.eps}
 \includegraphics[height=5cm,width=5cm]{Pp_Ma_variation_elep_1500MeV_sigma0.eps}
 \end{center}
 \caption{{$P_L(Q^2)$ and $P_P(Q^2)$ \textit{vs.} $Q^2$ for the  process $e^- p \to \nu_e \Sigma^0$ at $E_e^L$ = 0.5 GeV (left panel), $E_e^L$ = 1 GeV (central 
 panel) and $E_e^L$ = 1.5 GeV (right panel). The results are presented for different values of $M_A$ \textit{viz.} 0.9 GeV (solid line), 1.026 GeV 
 (dashed line), 1.1 GeV (dash-dotted line) and 1.2 GeV (dash--double-dotted line).}}\label{ma_var_pol_sig}
\end{figure*}
\begin{figure*}
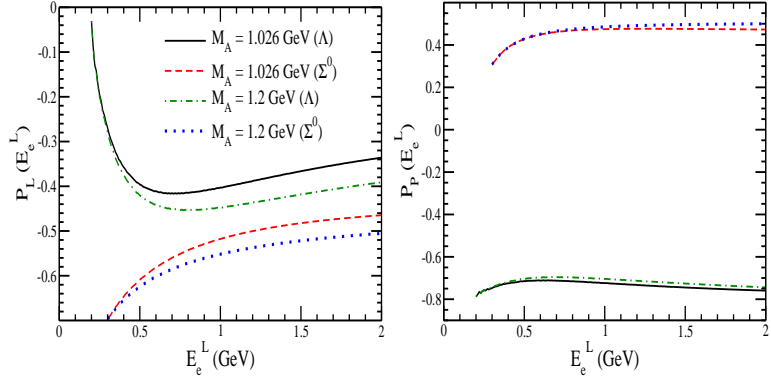

\begin{center}
 \includegraphics[height=5cm,width=5cm]{Plong_vs_elep.eps}
 \includegraphics[height=5cm,width=5cm]{Pperp_vs_elep.eps}
 \end{center}
 \caption{{ The results are presented for $P_L(E_e^L)$ (left panel) and $P_P(E_e^L)$ (right panel) \textit{vs.} $E_e^L$ using Eq.~(\ref{PLP_energy}) for the 
 process $e^- p \to \nu_e \Lambda$ at $M_A$ = 1.026 GeV (solid line) and 1.2 GeV (dash-dotted line), and
 for $e^- p \to \nu_e \Sigma^0$ at $M_A$ = 1.026 GeV (dashed line) and 1.2 GeV (dotted line).}}\label{energy_int}
\end{figure*}
\section{Results and discussion}\label{results}
We have used Eq.~(\ref{crosv.eq}) for the calculation of the total cross section $\sigma(E_e^L)$ and the
differential cross sections ($\frac{d\sigma}{dQ^2}$ and $\frac{d\sigma}{d\Omega}$), and Eqs. (\ref{sl}) and (\ref{st}) for the 
 longitudinal $P_L(Q^2)$ and perpendicular $P_P(Q^2)$ components of polarization, respectively, for the processes $e^- p \to \nu_e \Lambda$ and 
$e^- p \to \nu_e \Sigma^0$. The form factors are given in Table-\ref{tab:formfac}. For the vector nucleon form factors, we have used the 
parameterization of Bradford \textit{et al.}~\cite{Bradford:2006yz}. 
A dipole parameterization for the nucleon axial vector form factor with the dipole mass $M_A = 1.026 $ GeV~\cite{Bernard:2001rs} has been used. 
For the $\Delta^0$ production cross section, we have used Eq.~(\ref{delta_cross_section}) with the form factors defined in Eqs.~(\ref{civ_lala})
--(\ref{cia_lala})
and integrated over the angles to get the total cross section ($\sigma(\Delta)$).

    In Fig.~\ref{sigma1}, we have presented the results of $\sigma$ \textit{vs.} $E_e^L$
   for $\Lambda$, $\Sigma^0$ and $\Delta^0$ productions. In the inset of Fig.~\ref{sigma1},
  we have also presented the results for the ratio $R(E_e^L)=\frac{\sigma_Y}{\sigma_{\Delta^0}}$, for 
  Y = $\Lambda$ and $\Sigma^0$ productions. We observe that for energies $E_e^L~<~$0.4 GeV, the $\Lambda$ production cross section 
  is more than the $\Delta^0$ production which reduces to  $\sim 24\%$ of the $\Delta^0$ production for $E_e^L~\sim~$0.6 GeV and $\sim$ 16$\%$ for 
  $E_e^L =$ 1 GeV. 
 Thus, in the low electron energy range, the hyperons ($\Lambda,\Sigma^0$) give considerable contribution to the total cross section 
 along with the $\Delta^0$ production process. 
   The hyperon and $\Delta^0$ produced in these reactions decay to pion and nucleon. These particles 
 may be observed in coincidence. With the availability of the high luminosity electron beam
    (say 10$^{39}$/cm$^2$/s), we may be able to observe $\sim$ 665 events for the $\Delta^0$ production and $\sim$ 248 and 20 events for 
    $\Lambda$ and $\Sigma^0$ productions in the duration of 1 hour for 0.5 GeV electron energy, while almost the same number of events $\sim$ 150 
    for $\Lambda$ and $\Delta^0$ productions at $E_e^L~=~$0.4 GeV.
  
  To see the dependence of $\sigma$ on the axial dipole mass $M_A$, in Fig.~\ref{ma_sig}, 
  we have shown the results with $M_A$=0.9, 1.026, 1.1 and 1.2 GeV~\cite{Wilkinson:2016wmz,Ankowski:2016bji,Stowell:2016exm}.
  We find that the $\Lambda$ production cross section has larger sensitivity to 
  $M_A$ than the $\Sigma^0$ production cross section. It should be possible to determine the value of $M_A$ in the strangeness sector by observing the total $\Lambda$ 
  production in the energy range of 0.8$< E_e^L < $2 GeV.
  
   In Fig.~\ref{dsigma1}, we have presented the results for $\frac{d\sigma}{dQ^2}$ \textit{vs.} $Q^2$ at different values of the 
  electron energies \textit{viz.} $E_e^L$=0.4, 0.8 and 1.2 GeV, for $\Lambda$, $\Sigma^0$ and $\Delta^0$ productions. 
   In the threshold region, at very low $Q^2$, there is almost equal contribution from the $\Lambda$ and $\Delta^0$ productions.
     For $Q^2 >$ 0.1 GeV$^2$ there is a sharp fall in the $\Delta^0$ production cross section, whereas the $\Lambda$ production cross section decreases 
     slowly, similar to $e^- p \to \nu_e n$ reaction.
     At $E_e^L =$ 0.8 GeV, the $\Lambda$ cross section is $\sim$ 10--30 $\%$ of the $\Delta^0$ cross section in the low $Q^2$ region.

  In Fig.~\ref{dom1}, we have presented the results for the angular distribution $\frac{d\sigma}{d\Omega}$ 
 for $\Lambda$ and $\Sigma^0$ in $e^- p \to \nu_e \Lambda$ and $e^- p \to \nu_e \Sigma^0$ reactions
 at different electron energies $E_e^L$=0.5, 1, 2 and 5 GeV. In general,
  the nature of the angular distribution is qualitatively similar at these energies. However, the peak shifts towards the smaller angles at 
  lower $E_e^L$.
 We find that (not shown here) for $e^- p \to \nu_e \Lambda$ process, 
 the major contribution to the cross section comes from 
 $g_1^2(Q^2)$ and $f_1^2(Q^2)$ terms. Quantitatively, the contribution of $g_1^2(Q^2)$ is larger at the smaller angles while the contribution from $f_1^2(Q^2)$ is larger 
 in the peak region. The contributions of the interference terms like $f_1(Q^2)g_1(Q^2)$ and $f_2(Q^2)g_1(Q^2)$ are almost of the same strength. The 
 contribution from the $f_1(Q^2)f_2(Q^2)$ term
  is almost of equal strength at the smaller angles but becomes almost an order of magnitude smaller in the peak region as compared to the contribution
  of the vector-axial vector interference terms.
  For the process $e^- p \to \nu_e \Sigma^0$, it is the $f_2^2(Q^2)$ term which dominates at the smaller angles followed by the $g_1^2(Q^2)$ and 
  $f_1^2(Q^2)$ terms. However, in the peak region, the $f_1^2(Q^2)$ term dominates followed by the $f_2^2(Q^2)$ and $g_1^2(Q^2)$ terms. The term
  $f_2(Q^2)g_1(Q^2)$ is the dominant interference term. 
   We also find that there is not much effect of different parameterizations for the vector nucleon form factors $f^{p,n}_{1,2}$ available 
  in the literature on the angular distribution for both $\Lambda$ and $\Sigma^0$.
      The present results are in agreement with the results of Mintz and collaborators~\cite{Mintz:2004eu,Mintz:2002cj,Mintz:2001jc}.
   
The angular distribution $(\frac{d\sigma}{d\Omega})$ for $e^- p \to \nu_e \Lambda$ and $e^- p \to \nu_e \Sigma^0$  reactions have also been calculated 
 by Hwang \textit{et al.}~\cite{Hwang:1988fp} using two different models, \textit{i.e.}, the
Dirac Harmonic Oscillator Model (DHOM) and the MIT bag model, for calculating the form factors.  
In Fig.~\ref{hwang_com}, we have compared our results with the results obtained in these quark models at the incident electron energy $E_e^L = $ 0.5 GeV. 
Our results are qualitatively similar to their results but are 
quantitatively smaller specially in the case of $\Lambda$ production due to the different couplings used in the numerical calculations.

In Fig.~\ref{ma_dom}, the results are presented for $\frac{d\sigma}{d\Omega}$ for the processes  
 $e^- p \to \nu_e \Lambda$ and $e^- p \to \nu_e \Sigma^0$ by varying $M_A$ from 0.9 GeV to 1.2 GeV
 at the two incident electron energies of $E_e^L =$ 0.5 GeV and $E_e^L =$ 1 GeV. We find that the sensitivity of $\frac{d\sigma}{d\Omega}$ to the axial
 vector form factor is more for $e^- p \to \nu_e \Lambda$ than $e^- p \to \nu_e \Sigma^0$ process.
 It should be possible to determine the values of $M_A$ from the observation of $\frac{d\sigma}{d\Omega}$ for $e^- p \to \nu_e \Lambda$.

 In Figs.~\ref{ma_var_pol_lam}--\ref{energy_int}, we present the results for the longitudinal and perpendicular polarization observables.
 To study the dependence of $P_L(Q^2)$ and $P_P(Q^2)$ on $M_A$, we have presented the results for $P_L(Q^2)$ and $P_P(Q^2)$  at the incident electron
 energies $E_e^L=$ 0.5, 1 and 1.5 GeV for $e^- p \to \nu_e \Lambda$ process in Fig.~\ref{ma_var_pol_lam} 
 and $e^- p \to \nu_e \Sigma^0$ process in Fig.~\ref{ma_var_pol_sig}, respectively, by taking the different values of $M_A$, from 0.9 GeV to 1.2 GeV~\cite{Wilkinson:2016wmz,Ankowski:2016bji,Stowell:2016exm}. 
 We observe that the polarization observables ($P_L(Q^2)$ and $P_P(Q^2)$) in case of the $\Sigma^0$ production are more sensitive to the variation 
 in the value of $M_A$ as compared to the $\Lambda$ production.
 Also with the increase in energy, the sensitivity of the polarization observables especially $P_L(Q^2)$ increases for both $\Lambda$ and $\Sigma^0$
 which is clearly evident as the percentage difference in $P_L(Q^2)$ at $Q^2=$ 0.15 GeV$^2$ is $\sim$ 4$\%$(7$\%$) for $E_e^L=$ 0.5 GeV for 
 $\Lambda(\Sigma^0)$ and at $Q^2=$ 0.8 GeV$^2$ 
 is $\sim$ 2$\%$(7$\%$) for $E_e^L=$ 1 GeV and $\sim$ 6$\%$(28$\%$) for $E_e^L=$ 1.5 GeV for $\Lambda(\Sigma^0)$.

 To see the dependence of the polarization observables on $E_e^L$, we have integrated $P_L(Q^2)$ and $P_P(Q^2)$ over $Q^2$ 
  and obtained $P_{L}(E_e^L)$ and $P_{P}(E_e^L)$ defined as \cite{Akbar:2016awk}
\begin{equation}\label{PLP_energy}
 P_{L,P}(E_e^L) = \frac{\int^{Q^2_{max}}_{Q^2_{min}} P_{L,P} (Q^2) \frac{d\sigma}{dQ^2}dQ^2}{\int^{Q^2_{max}}_{Q^2_{min}}\frac{d\sigma}{dQ^2}dQ^2}.
\end{equation}
The results for the polarization components $P_{L}(E_e^L)$, $P_{P}(E_e^L)$ \textit{vs.} $E_e^L$ are presented in Fig.~\ref{energy_int} 
 for the $e^-p \to \nu_e \Lambda$ and $e^-p \to \nu_e \Sigma^0$ processes at the two
 different values of $M_A=$ 1.026 GeV and 1.2 GeV. From the figure, it may be observed that $P_L(E_e^L)$ is more sensitive to this variation in $M_A$ 
 than $P_P(E_e^L)$. 

\section{Summary}\label{summary}
We have studied in this work the differential and total scattering cross sections 
 as well as the longitudinal and perpendicular components of the polarization 
for $\Lambda$ and $\Sigma^0$ hyperons produced in the quasielastic reaction
of the electron on free proton. The form factors for the nucleon-hyperon transition have been obtained using the Cabibbo theory
assuming SU(3) invariance. 
 The sensitivity of $\frac{d\sigma}{d\Omega}$, $\sigma$, $P_L(Q^2)$ and $P_P(Q^2)$ to the axial mass $M_A$ has been studied. 
 
To summarize our results we find that:
  \begin{enumerate}
  \item[1)] Even though the production of the hyperons ($\Lambda,\Sigma^0$) is Cabibbo suppressed as compared to the $\Delta^0$ production, it may be 
  comparable to the $\Delta^0$ production
  in the region of low electron energies due to the threshold effects. We find that in the energy region of 0.4 to 0.8 GeV, 
  the $\Lambda$ production could be $\sim$ 80$\%$--17$\%$ of the $\Delta^0$ production. The cross sections are of the order of $10^{-41}$ cm$^2$ which 
  could be observed at the electron accelerators
  specifically at MAINZ and JLab with the low energy electron beams.

  \item[2)] We observe that the differential as well as the total cross section
  for the $\Lambda$ production is more sensitive to the variation in the value of $M_A$ 
 as compared to the $\Sigma^0$ case. This is because in the case of $\Lambda$ production
  the dominant contribution to the cross section comes from the axial vector form factor $g_1(Q^2)$, whereas the vector form factor $f_2(Q^2)$ 
  dominates in the case of $\Sigma^0$.
  
  \item[3)] The longitudinal and perpendicular components of polarization $P_L(Q^2)$ and $P_P(Q^2)$ are sensitive to the value of axial dipole mass $M_A$, 
  especially $P_L(Q^2)$ for $\Lambda$ 
  as well as $\Sigma^0$ production. It will enable us to make the measurements for the axial vector form factor independent of the cross section 
 measurements.

  \end{enumerate}
  \renewcommand{\theequation}{A-\arabic{equation}}
  \setcounter{equation}{0}  
  \section*{Appendix A}  
 
Expressions of $\alpha(Q^2)$ and $\beta(Q^2)$ in terms of Mandelstam variables:
\begin{eqnarray*}
 s &=& m_e^2 + m_N^2 + 2m_N E_e,\\
 t &=& -Q^2,\\
 u &=& m_e^2 + m_N^2 + m_Y^2 - s - t
\end{eqnarray*}
are
\begin{eqnarray}
 \alpha(Q^2) &=& 32\left[f_1^2(Q^2) \left((m_N+m_Y) \left((m_N-m_Y)^2-t\right) \right)\right. 
 + \frac{f_2^2(Q^2)}{(m_N+m_Y)^2}\left((m_N+m_Y)t \left((m_N-m_Y)^2-t\right) \right)\nonumber\\
 &+& g_1^2(Q^2) \left( (m_N-m_Y) \left(t-(m_N+m_Y)^2\right)\right) 
  + f_1(Q^2) g_1(Q^2)\left(-2 m_Y \left(m_N^2+2 m_e^2+m_Y^2-2 s-t\right) \right) \nonumber\\
 &+& \frac{f_1(Q^2) f_2(Q^2)}{(m_N+m_Y)} \left(\left(m_N^2-m_Y^2\right)^2-4 m_N m_Y t-t^2 \right) 
 + \frac{f_2(Q^2) g_1(Q^2)}{(m_N+m_Y)} \left(m_N^4+m_N^2 \left(m_e^2-2 (s+t)\right)\right.\nonumber\\
 &-&2 m_N m_e^2 m_Y -  m_e^2 \left(3 m_Y^2+t\right)  
 -\left.\left.\left(m_Y^2+t\right) \left(m_Y^2-2 s-t\right) \right)\right] ,
\end{eqnarray}

\begin{eqnarray}
 \beta(Q^2) &=& \frac{16}{m_Y} \left[f_1^2(Q^2) \left(-2 m_N^3 m_Y+m_N^2 \left(m_e^2+2 m_Y^2-t\right)\right.\right.  
 + \left. 2 m_N m_Y(s+t)+\left(m_Y^2-t\right) \left(m_e^2-2 s-t\right) \right) \nonumber\\
 &+& \frac{f_2^2(Q^2)}{(m_N+m_Y)^2}(m_N+m_Y) \left(m_e^2 (m_N-m_Y) \left(m_N^2+m_Y^2\right)\right. \nonumber\\
 &-&t \left(m_N^3+m_N^2 m_Y \right. + \left.\left. m_N \left(m_e^2-m_Y^2-2 s\right)\right.\right. 
 - \left.\left.m_e^2 m_Y+m_Y^3\right)+t^2 (m_N+m_Y) \right) \nonumber\\
 &+& g_1^2(Q^2) \left(2 m_N^3 m_Y+m_N^2 \left(m_e^2+2 m_Y^2-t\right)\right. 
 -2 \left. m_N m_Y (s+t) +\left(m_Y^2-t\right) \left(m_e^2-2 s-t\right)\right) \nonumber\\
 &+& f_1(Q^2) g_1(Q^2)\left(2 \left(m_N^2 \left(-m_e^2+2 s+t\right)\right.\right. 
 + \left.\left. m_e^2 \left(m_Y^2+2 s+t\right)+m_Y^2 t - 2 s^2-2 s t-t^2\right) \right)\nonumber\\
 &+& \frac{f_1(Q^2) f_2(Q^2)}{(m_N+m_Y)}\left(2 \left(m_N^4 (-m_Y)+m_N^3 \left(m_e^2-t\right)\right.\right. 
 + m_N^2 m_Y \left(m_Y^2+s\right)+m_N \left(m_e^2 \left(m_Y^2-t\right)\right.  \nonumber\\
 &+& \left. t \left(m_Y^2+2 s+t\right)\right)- \left.\left.m_Y \left(m_Y^2-t\right) (s+t)\right) \right)\nonumber\\
 &+& \frac{f_2(Q^2) g_1(Q^2)}{(m_N+m_Y)} \left(2 \left(m_N^4 (-m_Y)+m_N^3 \left(t-m_e^2\right)\right.\right. 
 - m_N^2 m_Y \left(m_e^2+m_Y^2-3 s-2 t\right) \nonumber\\
 &+& m_N \left(m_e^2 (s+t)+t \left(m_Y^2-t\right)\right) 
 + m_Y \left(m_e^2 \left(m_Y^2+2 s+t\right)\right. 
+ \left.\left.\left.\left.(s+t) \left(m_Y^2-2 s -t\right)\right)\right)\right)\right]
\end{eqnarray}

\end{document}